\documentclass[aps,pra,showpacs,superscriptaddress,twocolumn,amsmath,amssymb]{revtex4}
\usepackage[english]{babel}
\usepackage{latexsym}
\usepackage{graphics}
\usepackage{subfigure}
\usepackage{epsfig}

\begin{document}

\title{One-dimensional description of a Bose-Einstein condensate in a
rotating closed-loop waveguide}

\author{S. Schwartz}
\affiliation{Thales Research and Technology France, RD 128, F-91767 Palaiseau Cedex, France}
\affiliation{Laboratoire Charles Fabry de l'Institut
d'Optique, UMR8501 du CNRS, Centre scientifique d'Orsay B\^at. 503, 91403 Orsay Cedex, France}
\affiliation{BEC-CNR-INFM and Dipartimento di Fisica,
Universit\`a di Trento, I-38050 Povo, Italy}

\author{M. Cozzini}
\affiliation{ISI Foundation, Villa Gualino, I-10133 Torino, Italy}

\author{C. Menotti}
\affiliation{BEC-CNR-INFM and Dipartimento di
Fisica, Universit\`a di Trento, I-38050 Povo, Italy}
\affiliation{ICFO - The Instute for Photonic Sciences,
 Mediterranean Technology Park, \\
 E-08860 Castelldefels (Barcelona), Spain}

\author{I. Carusotto} \affiliation{BEC-CNR-INFM and Dipartimento di
Fisica, Universit\`a di Trento, I-38050 Povo, Italy}

\author{P. Bouyer}
\affiliation{Laboratoire Charles Fabry de l'Institut d'Optique, UMR8501 du CNRS, Centre scientifique d'Orsay B\^at. 503, 91403 Orsay Cedex, France}

\author{S. Stringari}
\affiliation{BEC-CNR-INFM and Dipartimento di Fisica, Universit\`a di Trento,
I-38050 Povo, Italy}

\date{\today}

\begin{abstract}
We propose a general procedure for reducing the three-dimensional
Schr\"odinger equation for atoms moving along a strongly confining
atomic waveguide to an effective one-dimensional equation. This
procedure is applied to the case of a rotating closed-loop
waveguide. The possibility of including mean-field atomic interactions
is presented. Application of the general theory to characterize a new
concept of atomic waveguide based on optical tweezers is finally
discussed.
\end{abstract}

\maketitle

\section{Introduction}

A very promising and challenging experiment to be performed in the
near future using coherent matter waves is the observation of a
rotation-induced supercurrent around a closed loop.  This not only
will attract a broad interest from the point of view of fundamental
physics as a direct manifestation of superfluidity~\cite{Leggett}, but
may also open the way to new kinds of high-precision rotation sensors
based on matter-wave~\cite{AtomicGyro}, rather than
light-wave~\cite{LightGyro} interferometry.  The irrotationality
constraint of superfluids is in fact softened in the multiply
connected geometry of a closed loop waveguide, which then appears as
an ideal playground for the study of quantized vorticity and the
related quantum interference phenomena. 

As it has been originally pointed out by Bloch \cite{Bloch}, and then
developed in more detail by Ueda and co-workers \cite{Ueda}, an
appropriately tailored rotating potential can be used to transfer
vorticity into a Bose-Einstein condensate trapped in a closed loop
geometry.  The 2D numerical simulations reported in~\cite{Ueda}, which
fully include the effect of interactions, have demonstrated a close
qualitative analogy between the physics of rotating condensates and
the physics of condensates in 1D optical lattices, an analogy that can
be made quantitative by correctly taking into account the
modifications to the trapping potential due to rotation. 
Metastability appears for a rotating flux as a consequence of flux
quantization: points of vanishing density have in fact to be
introduced in the wavefunction if quantized vortices are to penetrate
or exit the cloud \cite{Leggett}.  When the velocity of the superfluid
flow exceeds the velocity of sound, the system becomes energetically
unstable according to the Landau criterion of superfluidity.  In this
regime, the interaction of the flowing fluid with a stationary defect
is able to create excitations in the fluid, which can eventually lead
to dissipation of the supercurrent by means of phase slippage
processes~\cite{Ueda2,Pavloff}.

In order to simplify the theoretical study of the generation and
dynamics of supercurrents in realistic loop configurations (e.g. a
toroidal trap), it is very convenient to be able to isolate the
longitudinal dynamics along the, possibly rotating, waveguide so to
reduce the full 3D problem to 1D. This is one of
the central points of the present paper.

Once the 3D problem is reduced to a 1D one, the formal connection
between annular rotating Bose-Einstein condensates and Bose-Einstein
condensates in optical lattices becomes apparent, and one can
start taking fully advantage of the large amount of literature
appeared on  the latter subject.
A remarkable example of this connection are the so-called swallow
tails (instable parts of the band structure at the edge and in the centre of
the Brillouin zone) shown by the band structure of a BEC flowing in an
optical lattice \cite{Mueller, Pethick}: as it has been pointed out
in Ref.~\cite{Ueda}, they play an important role in the nucleation of
vortices and supercurrents in rotating BECs.
Further very interesting phenomena that have been studied in the
context of BECs in optical lattices are dynamical instabilities
\cite{Niu, Menotti, Paraoanu, Inguscio} and gap solitons
\cite{Oberthaler}, which are expected to have interesting counterparts
in the physics of rotating BECs.

First observations~\cite{StorageRings} of Bose-Einstein condensates trapped in toroidal
traps have been recently reported in Refs.~\cite{Stamper_Kurn,Riis}.
These experimental setups were based on magnetic potential, which
introduces some limitations on the geometries that can be obtained and, more
specifically, on the flexibility of the setup with respect to time and space
modulations of the confining potential.

The other central point of our paper is in fact to propose a
novel realization of toroidal atomic waveguide, that makes
use of the attractive optical potential of a red-detuned laser beam as
an optical tweezer~\cite{tweezer}. 
The main advantage of our proposal with respect to recent related
ones~\cite{TORT_th} is the flexibility of the setup.
A toroidal trapping potential with a strong transverse confinement
is obtained by rapidly moving the focus point of the laser.  As the 
trajectory of the focus point can be arbitrarily chosen, as well as
its speed of motion, any shape can in principle be obtained for the
1D waveguide, and any longitudinal potential can be applied onto the
atoms in addition to the transverse confinement.
Furthermore, the shape of the curve can be changed in time, so to
obtain, e.g., rotating waveguide traps.

The paper is organized as follows.  In Sec.~\ref{quantitative} we put
the problem on a precise mathematical basis and then in
Sec.~\ref{decoupling} we introduce our decoupling scheme to reduce the
three-dimensional Schr\"odinger equation to an effective
one-dimensional one.  Sec.~\ref{longitudinal} and
Sec.~\ref{interactions} discuss the effect of a longitudinal
dependence of the transverse trapping frequency, and the effect of the
interatomic interactions.  The main theoretical results of this paper
are given in Sec.~\ref{rotating}, where the decoupling scheme is
generalized to the case of a rotating waveguide.  In
Sec.~\ref{experimental}, the theoretical approach is applied to our
proposal of 1D waveguide based on a rapidly moving optical tweezer.
Conclusions are finally drawn in Sec.~\ref{Conclu}.

\section{Quantitative description of the atomic waveguide}
\label{quantitative}

Consider an atomic waveguide whose axis follows a regular curve $\mathcal{C}$
parametrically defined by the vector $\mathbf{r}_\mathcal{C}(s)$, $s$
being the arclength coordinate along $\mathcal{C}$.  At each point of the
curve, we can define the Frenet frame $(\mathbf{t}, \mathbf{n}, \mathbf{b})$
as

\begin{eqnarray}
\mathbf{t} &=&\frac{\textrm{d}\mathbf{r}_\mathcal{C}}{\textrm{d} s } \;, \\
\kappa\, \mathbf{n}  &=& \frac{\textrm{d}\mathbf{t}}{\textrm{d} s } \;,\\
\tau\, \mathbf{b} &=&\frac{\textrm{d}\mathbf{n}}{\textrm{d} s }+ \kappa \mathbf{t} \;, \\
\frac{\textrm{d}\mathbf{b}}{\textrm{d}s}&=& -\tau \mathbf{n} \;,
\end{eqnarray}
where $\kappa$ and $\tau$ are respectively known as the curvature and the torsion of
$\mathcal{C}$~\cite{DoCarmo}.  In the vicinity of $\mathcal{C}$, a
local system $(s,u,v)$ of coordinates can be introduced, such that

\begin{equation}
\mathbf{r} ( s , u , v ) = \mathbf{r}_\mathcal{C} (s) + u
\mathbf{N}(s)
+ v \mathbf{B}(s). \label{suv}
\end{equation}
The transverse frame $(\mathbf{t},\mathbf{N},\mathbf{B})$ is related
to the Frenet frame $(\mathbf{t},\mathbf{n},\mathbf{b})$ by a simple rotation
around ${\mathbf t}$

\begin{equation} \nonumber
\left( \begin{array}{c} \mathbf{N} \\ \mathbf{B} \end{array} \right) =
\left( \begin{array}{cc} \cos \theta & \sin \theta \\ -\sin \theta &
\cos \theta
\end{array} \right) \left( \begin{array}{c} \mathbf{n} \\ \mathbf{b} \end{array} \right)
\end{equation}
of an angle $\theta$ such that

\begin{equation}
\frac{\textrm{d} \theta}{\textrm{d} s}=-\tau\,.
\end{equation}
With this choice of coordinates, the gradient has the following simple form

\begin{equation}
\nabla = \mathbf{t}\,h^{-1}\,\partial_s +\mathbf{N}\,\partial_u
+\mathbf{B}\,\partial_v
\label{eq:gradient}
\end{equation}
where the scale factor

\begin{equation}
h(s,u,v) = 1- \kappa \left(u \cos \theta - v \sin\theta\right)
\label{h}
\end{equation}
depends on the torsion $\tau$ only via the angle $\theta$.

The transverse confinement in the $(\mathbf{N},\mathbf{B})$ plane
orthogonal to the waveguide axis is assumed to be harmonic and of the form

\begin{equation} \label{tranchee}
V_\perp(u,v) = \frac{M}{2}\,\left(\omega_{u}^2\, u^2 +
\omega_{v}^2\,v^2 \right),
\end{equation}
where $M$ is the atomic mass and $\omega_{u,v}$ are the transverse
trapping frequencies in respectively the $\mathbf{N}$ and $\mathbf{B}$
directions (which can depend on the longitudinal coordinate $s$).  For
the sake of simplicity, the discussion that follows will be restricted
to two mostly significant cases.  In Sec.~\ref{decoupling}, the curve
is allowed to have a non-vanishing torsion $\tau$, but the transverse
trapping is taken as isotropic $\omega_u=\omega_v=\omega_\perp$.  This
condition is enough to rule out effects due to the finite
torsion~\cite{Pavloff}.  In the Secs.\ref{rotating} and
\ref{experimental} a different situation is considered, where the
curve is taken to belong to the plane orthogonal to the rotation axis,
while the trapping potential can have different frequencies
$\omega_u\neq \omega_v$ in the two orthogonal directions respectively
in and perpendicular to the plane.

The key assumption of our treatment is the strong confinement
hypothesis, where the extension

\begin{equation}
\sigma = \sqrt{\frac{\hbar}{M \omega_\perp}}
\end{equation}
of the transverse ground state is much smaller than all typical length
scales of the curve $\mathcal{C}$, namely

\begin{equation}\label{adiabatic}
\kappa \sigma \ll 1 \qquad |\kappa'| \sigma \ll |\kappa| \qquad |\kappa''| \sigma \ll \kappa^2 \;.
\end{equation}
Here, and throughout the whole paper, primed quantities denote derivation with
respect to the longitudinal coordinate $s$. The reason why conditions
(\ref{adiabatic}) involve up to the second derivatives of the curvature will
become clear in the following. These conditions also guarantee that the
coordinate system (\ref{suv}) can be safely used to describe the transverse
extension of the wavefunction.

If the waveguide is at rest in a reference frame rotating at an angular
speed $\mathbf{\Omega}$, the stationary Schr\"odinger equation in the rotating
waveguide has then the form~\cite{Landau,CCT_CdF}

\begin{equation} \label{schrodinger}
\begin{split}
 \mu \Psi (s,u,v) & = \bigg[ -\frac{\hbar^2}{2M}\,\nabla^2 -
\mathbf{\Omega} \cdot (\mathbf{r}\times \mathbf{p}) \\ & +
V_\perp(u,v) +V_{\rm ext}(s) \bigg] \Psi (s,u,v) \;,
\end{split}
\end{equation}
where the momentum operator is defined as usual as $\mathbf{p} = -i
\hbar \nabla$.  In the $(s,u,v)$ coordinates, the gradient has the
form (\ref{eq:gradient}). The term $V_{\rm ext}(s)$ describes any weak
potential acting on the atoms along the direction of the waveguide in
addition to the waveguide trapping.

Note that the normalization of the wavefunction $\Psi$ has to take into
account the new metric associated to the $(s,u,v)$ coordinates
\begin{equation}
\int \!\!\!\!\int \!\!\!\!\int \textrm{d} s \,\textrm{d} u
 \,\textrm{d} v \, h(s,u,v) \, |\Psi( s,u,v )|^2 = 1 \;.
\end{equation}
As $h(s,u,v)$, defined in (\ref{h}), is not factorizable as a product
of functions of respectively the longitudinal $s$ and transverse $(u,v)$
coordinates, this condition is not convenient for decoupling the
transverse and the longitudinal dynamics.  It is then useful to introduce a rescaled
wavefunction $\Phi=\sqrt{h} \Psi$ whose normalization is the usual one
\begin{equation} \label{normalization}
\int\!\!\!\!\int \!\!\!\!\int \textrm{d} s \,\textrm{d} u \,\textrm{d} v \, |\Phi( s,u,v )|^2 = 1\,.
\end{equation}
The Hamiltonian operator acting on the wavefunction $\Phi(s,u,v)$ then
has the form

\begin{equation}
\begin{split}
\hat{H} &= -\frac{\hbar^2} {2M} \bigg[ \partial_s \left(h^{-2}(s,u,v)
\partial_s \right)+ \partial^2_{uu} + \partial^2_{vv} \\& + \frac{
\kappa^2(s)}{4h^2(s,u,v)} + \frac{5[h'(s,u,v)]^2}{4h^4(s,u,v)} -
\frac{h''(s,u,v)}{2h^3(s,u,v)} \bigg] \\& - \sqrt{h(s,u,v)} \,
\mathbf{\Omega} \cdot (\mathbf{r}\times
\mathbf{p})\frac{1}{\sqrt{h(s,u,v)}} \\ & + V_\perp(u,v) +V_{\rm
ext}(s) \;.
\end{split}
\end{equation}
In the next sections, we shall proceed with the decoupling of the
transverse and the longitudinal dynamics under the strong confinement
hypothesis.

\section{Decoupling procedure in the non-rotating case}
\label{decoupling}

In this section, we shall start by considering the simplest case of a
non-rotating waveguide ($\mathbf{\Omega}=0$) with a spatially constant
trapping frequency ($\omega_\perp'=0$).  It is useful to rewrite the
wavefunction $\Phi$ as the product $\Phi(s,u,v)=f_s(u,v)\,g(s) $ of a
longitudinal wavefunction $g(s)$ and a transverse wavefunction $f_s(u,v)$
in general dependent on the longitudinal coordinate $s$. The normalization
conditions can then be written as

\begin{equation}
\left\{
\begin{split}
& \int\!\!\!\!\int \textrm{d}u \, \textrm{d}v | f_{ s}(u,v )|^2 = 1,
\\ & \int\! \textrm{d}  s  |g( s )|^2 = 1 \,.
\end{split} \right.
\end{equation}
We now proceed in the spirit of the Born-Oppenheimer approximation
\cite{Born_Oppenheimer} where the fast electronic degrees of freedom
are eliminated and summarized as an effective potential acting on the
nuclei.

For any longitudinal wavefunction $g(s)$, we define a transverse Hamiltonian
$\hat{h}$ at the position $s$ such that

\begin{equation} \label{small_h}
\hat{h} f_s \equiv \hat{H} (f_s g).
\end{equation}
The key assumption of our approach is to assume the transverse motion to be
frozen in the ground state $f_s^0$ of $\hat{h}$.
Our aim is to reduce the three-dimensional problem to a one-dimensional one,
by integrating over the transverse degrees of freedom

\begin{equation}
\label{Schrodinger projected}
\mu g(s) =\int \!\!\!\!\int \textrm{d} u \,\textrm{d} v \, f_s^{0*}(u,v) \hat{H} f_s^0(u,v) g(s)
\equiv \hat{H}_g g(s),
\end{equation}
hence eliminating adiabatically the transverse motion. An explicit form of $\hat{H}_g$ can be obtained by means of a perturbative expansion by
separating in $\hat{h}$ the different contributions due to the transverse and longitudinal degrees of freedom, namely

\begin{equation} \label{eq f}
\hat{h} = H_0 + W  \;,
\end{equation}
with

\begin{equation}
H_0 = g(s) \left[ -\frac{\hbar^2} {2M} \left( \partial^2_{uu} +
\partial^2_{vv} \right) + V_\perp(u,v) \right],
\label{eq:H0}
\end{equation}
and

\begin{equation} \label{eq W}
\begin{split}
W & = -\frac{\hbar^2} {2M} \bigg[ h^{-2}(s,u,v) \big[g(s)
  \partial^2_{ss}+2 g'(s) \partial_s \\ & + g''(s) \big] - \frac{2h'(s,u,v)}{h^3(s,u,v)}
  \big[g(s) \partial_s+ g'(s) \big] \\ & + \frac{g(s) \kappa^2(s)}{4h^2(s,u,v)} + \frac{5 g(s)
  [h'(s,u,v)]^2}{4h^4(s,u,v)} \\ & - \frac{g(s) h''(s,u,v)}{2h^3(s,u,v)} \bigg] +g(s) V_{\rm ext}(s)  \;.
\end{split}
\end{equation}
We treat perturbatively the Hamiltonian $W$ with respect to $H_0$,
which a part from the multiplying factor $g$,
corresponds to an harmonic oscillator with energy $\hbar \omega_\perp$.
All terms of $W$ are much smaller than $|g|\hbar \omega_\perp$ provided

\begin{equation} \label{low_kinetic}
\left| g''\right| \ll |g| / \sigma^2 \quad \textrm{and} \quad \left|
g' \right| \ll |g| / \sigma \;,
\end{equation}
conditions which will be self-consistently verified at the end of the procedure,
thanks to conditions (\ref{adiabatic}).

Perturbation theory at first-order (in the small parameter $\kappa^2 \sigma^2$)
allows to replace $f_s^0$ in Eq.~(\ref{Schrodinger projected}) with
the ground state wavefunction $f_0$ of the harmonic oscillator of
frequency $\omega_\perp$, given by

\begin{equation}
f_0 (u,v) = \frac{1}{\sqrt{\pi} \sigma} e^{- (u^2+v^2)/(2\sigma^2)} \;,
\end{equation}
leading to the following effective one-dimensional
Schr\"odinger equation for the longitudinal wavefunction $g(s)$
\begin{equation} \label{1D_non_rotating}
\mu g = -\frac{\hbar^2} {2M} \frac{\textrm{d}^2 g}{\textrm{d} s ^2} +
 \left[\hbar \omega_{\perp}+V_{\rm ext}(s)- \frac{\hbar^2
 \kappa^2(s)}{8M}\right] g.
\end{equation}
We easily recognize the usual kinetic energy term, the zero-point
energy of the two-dimensional transverse trapping potential, and the
weak external potential $V_{\rm ext}$.
The last term of equation (\ref{1D_non_rotating}) gives an effective
potential proportional to the square of the curvature as discussed
in~\cite{Pavloff}.  The smoothness of the waveguide, as quantified by
conditions (\ref{adiabatic}),  guarantees that conditions (\ref{low_kinetic})
are satisfied and hence the self-consistency of the approach.

\begin{figure}
\begin{center}
\includegraphics[scale=0.46]{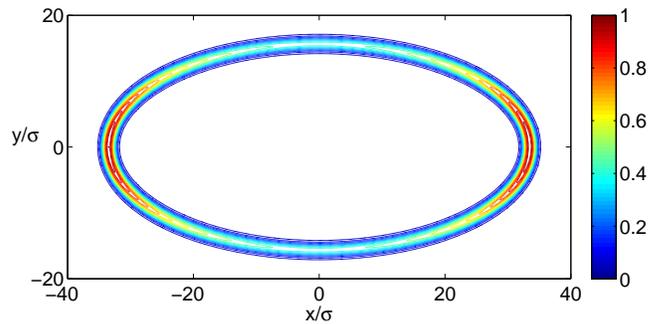}
\end{center}
\caption{Result of the numerical integration of the 2D Schr\"odinger
  equation in the case of an elliptical waveguide with strong
  transverse confinement.
Parameters: semiaxis $b/\sigma \simeq 34.5$; $a/\sigma \simeq 15.6$.
 \label{fig1}}
\end{figure}

This result can be illustrated on the specific example of an elliptical
waveguide, whose axis are respectively equal to $R \cosh \eta_0$ and $R \sinh
\eta_0$.  The parameter $\eta_0$ characterizes its eccentricity (the
larger $\eta_0$, the closer to a circle) and $R$ gives the overall scale.  The
strong confinement hypothesis is then $\sigma \ll R \sinh \eta_0$.
The one-dimension equation is

\begin{equation} \label{1D non rotating bis}
\mu g(s) = \left( -\frac{\hbar^2} {2M}
 \frac{\textrm{d}^2 }{\textrm{d} s^2} + \hbar \omega_{\perp} - \frac{\hbar^2 \kappa^2(s)} {8M} \right) g(s) \;.
\end{equation}
The curvature $\kappa$ has the simple expression

\begin{equation}
\kappa = \frac{\cosh \eta_0 \sinh \eta_0}{R (\sinh^2 \eta_0 + \sin^2 w)^{3/2}}\;,
\end{equation}
where $w$ is the parametric angle along the ellipse, related to the arclength
coordinate $s$ by the differential relation

\begin{equation}
\textrm{d} s = R \sqrt{\sinh^2 \eta_0 + \sin^2 w} \; \textrm{d} w \;.
\end{equation}
The curvature $\kappa$ has its maxima on the great axis of the ellipse.
The curvature-induced effective potential is thus minimum at these points. Its
effect is illustrated in Fig.\ref{fig1}, where the results of a numerical
integration of the 2D Schr\"odinger equation in cartesian coordinates are
shown (the third dimension was neglected for the sake of simplicity).
Due to the curvature-induced potential, the atomic
density is maximum on the great axis of the ellipse.

\begin{figure}
\begin{center}
\includegraphics[scale=0.5]{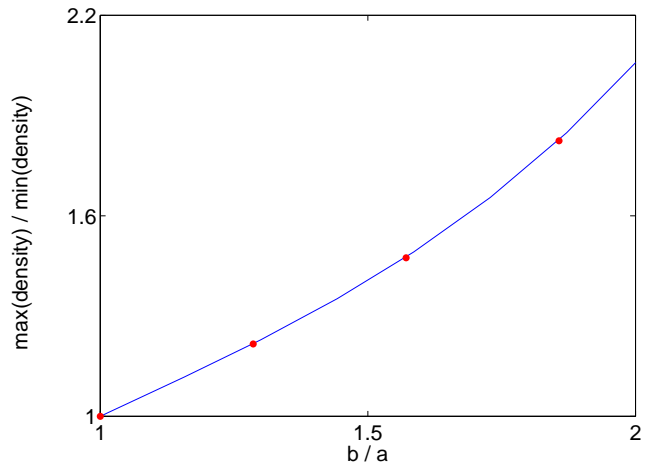}
\end{center}
\caption{Result of the numerical integration of the Schr\"odinger
  equation in the case of an elliptical waveguide with strong
  transverse confinement.  The graph shows the ratio between the
  density at the extrema of the two orthogonal axes of the ellipse, as
  a function of   the ratio between the length of theses axes
  $b/a$. The points come from the
  simulation of the 2D Schr\"odinger equation in cartesian
  coordinates, while the full line comes from the 1D simulation of
  equation~(\ref{1D_non_rotating}).}\label{fig:couplings}
\end{figure}

A more quantitative comparison between the full 2D Schr\"odinger
equation in cartesian coordinates and the effective 1D Schr\"odinger
equation~(\ref{1D_non_rotating}) is shown in Fig.\ref{fig:couplings}
for the case of an elliptical waveguide with strong
confinement. The agreement is remarkable.

\subsection{Effect of a longitudinal variation of $\omega_\perp$}
\label{longitudinal}

The case of a transverse trapping frequency $\omega_\perp(s)$ with
a non-trivial dependence on the longitudinal coordinate $s$ is addressed in the
present section.  This induces a  longitudinal variation of the transverse
wavefunction $f_{0s}$ and  introduce new terms in
$\hat{H}_g$  due to the non-vanishing longitudinal derivatives
of $f_{0s}$. Applying the same procedure as in the previous section, and
limiting ourselves to the first order in $\kappa^2 \sigma^2$,
we get the following 1D effective equation for $g(s)$

\begin{equation}
\mu\,g = - \frac{\hbar^2} {2M} \frac{\textrm{d}^2 g}{\textrm{d} s^2}
+\left[ \hbar \omega_{\perp}+V_{\rm ext}-\frac{\hbar^2 \kappa^2}
  {8M}
  +\frac{\hbar^2}{16M}\frac{\omega_\perp'^2}{\omega_\perp^2}\right]\, g.
\label{modul_omega_perp}
\end{equation}
The dependence of $\omega_\perp$ on $s$ not only gives an
$s$-dependent potential energy equal to zero-point energy
$\hbar\omega_\perp (s)$, but also adds a further contribution
proportional to $\omega_\perp'^2/\omega_\perp^2$.  Consistency with
our decoupling procedure requires that
$|\omega_\perp'|\lessapprox\kappa\,\omega_\perp$ and
$|\omega''_{\perp}| \lessapprox \kappa^2 \omega_\perp$.  This implies
in particular that the term proportional to
$\omega_\perp'^2/\omega_\perp^2$ is much smaller than the zero-point
energy term.

\begin{figure}
\begin{center}
\includegraphics[scale=0.47]{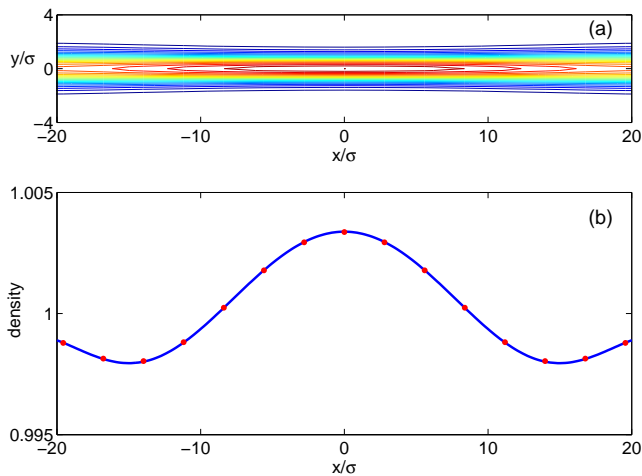}
\end{center}
\caption{(a) Result of the numerical integration of the 2D
  Schr\"odinger equation in the case of an straight waveguide with a
  strong confinement of spatially-dependent transverse confinement
  frequency $\omega_\perp(x)=\omega_\perp \sqrt{1+4(\delta
  \omega_\perp/\omega_\perp)\, \cos(\pi x/\lambda)}$.  The colour code
  is the same as in Fig.\ref{fig1}. Parameters: $\delta
  \omega_\perp/\omega_\perp=0.1$, $\lambda /\sigma \simeq 22.4$.  An
  external potential such that $V_{\rm ext}+V_\perp= M
  \omega_\perp(x)^2 y^2 /2 - \hbar \omega_\perp(x)/2$ has been
  introduced in order to compensate the spatial modulation of the
  zero-point trapping energy.  (b) comparison between the 2D density
  integrated along direction y (red dots) and the solution if the
  effective 1D Schr\"odinger equation (full line).}
  \label{fig_tube}
\end{figure}

In order to check the effect of the new potential term proportional to
$\omega_\perp'^2/\omega_\perp^2$, we have numerically solved the 2D
Schr\"odinger equation in cartesian coordinates for the case of a
straight waveguide with a modulated $\omega_\perp(s)$.  In order to
put the effect of the $\omega_\perp'^2/\omega_\perp^2$ in better
evidence, the external potential $V_{\rm ext}$ has been chosen in such
a way to compensate the modulation of the zero-point energy of the
transverse harmonic trapping $\hbar\omega_\perp(s)/2+V_{\rm
ext}(s)=0$. As shown in Fig.\ref{fig_tube}, the density modulation has
indeed the same periodicity of $\omega_\perp'^2/\omega_\perp^2$ and is
in quantitatively good agreement with the numerical solution of the
one-dimensional equation (\ref{modul_omega_perp}).

\subsection{Effect of interatomic interactions}
\label{interactions}

In a 3D geometry, mean-field interactions are included in the the
Gross-Pitaevskii equation~\cite{book} as nonlinear terms of the form:
\begin{equation} \label{Vint}
V_{int} = g_{3D} N_0 \left|\frac{f_s(u,v)g(s)}{\sqrt{h(s,u,v)}}\right|^2 \;,
\end{equation}
being $g_{3D}=4\pi\hbar^2a_{3D}/M$, $a_{3D}$ the three-dimensional
scattering length and $N_0$ is the total number of atoms
\cite{notation}.  The Gross-Pitaevskii equation in presence of a
periodic potential has been extensively studied in the context of
Bose-Einstein condensates in optical lattices. The issue of the
factorization of the wavefunction in its transverse and longitudinal
part in the presence of interactions is in general a non trivial one.

In the language of the present paper, the presence of interactions
requires that the transverse wavefunction $f_0(u,v)$ is now solution
of the ground state Gross-Pitaevskii equation
\begin{multline}
\Big[ -\frac{\hbar^2}{2m}
    (\partial^2_{uu} + \partial^2_{vv})+ N_0\,g_{3D}\,|g(s)|^2 \,
    |f_{0s}(u,v)|^2 + \\
+ V_\perp (u,v) \Big] f_{0s}(u,v)=
\mu_\perp \,f_{0s}(u,v)
\label{with_int}
\end{multline}
which results from the inclusion of the mean-field interaction term
into the transverse Hamiltonian (\ref{eq:H0}).
Here $\mu_\perp$ has the meaning of a transverse chemical potential.
In the general case, Eq.~(\ref{with_int}) has to be solved
self-consistently with the equation determining the longitudinal
wavefunction $g(s)$, which can be a computationally intensive task
unless clever schemes are adopted, as e.g. in~\cite{NPGPE}.

A very simple formulation can be obtained in the limiting case of very
strong radial confinement, when interactions have a negligible effect
on the shape of the transverse ground state wavefunction $f_0(u,v)$.
In this limit, interactions only provide a mean-field energy term
proportional to the local density to be included in the longitudinal
equation. This has then the usual form of a 1D Gross-Pitaevskii
equation including the curvature term discussed in the previous
section
\begin{multline}
 (\mu -\hbar \omega_\perp) g(s) = -\frac{\hbar^2} {2M}
  \frac{\textrm{d}^2 g(s)}{\textrm{d} s ^2} + V_{\rm ext}(s) g(s) \\ 
- \frac{\hbar^2 \kappa^2(s)}{8M}g(s) + g_{1D} N_0 |g(s)|^2 g(s),
\label{eq:gpe}
\end{multline}
where the effective 1D coupling constant is defined as:
\begin{equation}
g_\textrm{1D} = \frac{4 a_{3D} \hbar^2}{M \sigma^2} \;.
\end{equation}

\section{Case of a rotating waveguide}
\label{rotating}

We now turn to the more general case of a rotating waveguide. For simplicity,
we shall from now on assume that the curve $\mathcal{C}$ is included in the
plane $(x,y)$, that the rotation vector $\mathbf{\Omega}$ is orthogonal to
this plane, and that the origin of $\mathbf{r}_\mathcal{C}$ coincides with the
centre of rotation. We shall furthermore consider in what follows the case
of a constant $\omega_\perp$ independent of the position on the curve.

Repeating the same steps as in Sec.~\ref{decoupling}, the transverse
Hamiltonian can be decomposed as

\begin{equation}
\hat{h} = \tilde{H}_0 + \tilde{W} \;,
\end{equation}
with
\begin{equation}
\tilde{H}_0 = \frac{\hbar^2 g} {2M} \left(i \partial_u -\frac{M\Omega r_t}{\hbar} \right)^2 - \frac{\hbar^2 g} {2M} \partial^2_{vv} + g V_\perp \;,
\end{equation}
and

\begin{multline}
\tilde{W} = W + i\hbar \Omega \bigg[ \frac{g r_t \kappa }{2 h } + \frac{u}{h} g' - \frac{g u h'}{2h^2} + r_n g' \\ - \frac{g r_n h'}{2h^2} + \frac{g
u}{h} \partial_s + \frac{g r_t}{h} \partial_s \bigg] - \frac{1}{2} M \Omega^2 r_t^2 g \;.
\end{multline}
In the last expression, the following shorthand notations have been used:
$r_t = \mathbf{r}_\mathcal{C} \cdot \mathbf{t}$ and $r_n =
\mathbf{r}_\mathcal{C} \cdot \mathbf{n}$.  The ground state of $\tilde{H}_0$ is now given by

\begin{equation}
\tilde{f}_{0s} (u,v) = f_0(u,v)\, e^{-i M\Omega r_t u / \hbar} \;.
\end{equation}

Assuming a moderate rotation speed $|\Omega|\ll \omega_\perp$ and $M \Omega^2 \mathbf{r}_\mathcal{C}^2 \ll \hbar \omega_\perp$, an explicit calculation
of the longitudinal derivatives of $\tilde{f}_{0s}$ shows that the following inequalities are satisfied

\begin{equation}
\left| \frac{\partial \tilde{f}_{0s}}{\partial s} \right| \ll
\left|\frac{\tilde{f}_{0s}}{\sigma} \right| \qquad \textrm{and} \qquad
\left|
\frac{\partial^2 \tilde{f}_{0s}}{\partial s^2} \right| \ll
\left|\frac{\tilde{f}_{0s}}{\sigma^2}\right|,
\end{equation}
which guarantee that $\tilde{W}$ can be treated as a small
perturbation with respect to $\tilde{H}_0$.
To the first order in $\kappa^2 \sigma^2$, one then has

\begin{multline}
\hat{H}_g g(s)=\int\!\int\! \textrm{d}u\,\textrm{d}v
\,\tilde{f}_{0s}^{*} (\tilde{H}_0 + \tilde{W}) \tilde{f}_{0s}= \\ 
= - \frac{\hbar^2}{2M} g'' +\left[\hbar \omega_\perp+ V_{\rm ext}- \frac{
\hbar^2 \kappa^2}{8M} - \frac{M \Omega^2}{2} r_t^2\right]\,g \\ +i
\hbar \Omega \left[ \frac{r_t \kappa }{2}\,g + r_n \,g' \right]
\end{multline}
and the 1D equation for the rotating waveguide can finally be written as

\begin{multline}
\label{1D rotating}
\mu\, g = \frac{1} {2M} \left[ i\hbar \frac{\textrm{d}}{\textrm{d}s}
+M\, \mathbf{t} \cdot \left(\mathbf{r}_C \times
\mathbf{\Omega}\right) \right]^2 g \\ +\left[ \hbar \omega_{\perp}
+ V_{\rm ext}-\frac{\hbar^2 \kappa^2} {8M} - \frac{1}{2} M
\Omega^2 \mathbf{r}_C^2\right] g \;,
\end{multline}
where we have used the identity $ \textrm{d} r_n / \textrm{d} s =
\kappa r_t$.  As compared to the non-rotating case of
(\ref{1D_non_rotating}),
additional terms appear here due to the rotation. The first one is the
one-dimensional form of the gauge term appearing in the kinetic energy
term in the rotating frame and mostly affects the phase of the
wavefunction. The second rotation-induced term is the classical
centrifugal energy term.
This term, along with the curvature-induced term, the external potential term,
and the zero-point energy term, can be used to transfer angular momentum from the
trap to the atoms and then establish a finite phase circulation in the
condensate.

\subsection{Analogy with optical lattices}

Consider for simplicity a circular waveguide of radius $R$ in the
  presence of a rotating periodic potential of period $2\pi/\ell$
  ($\ell$ is integer) in the angular coordinate $\theta$

\begin{equation}
V(\theta,t)=V_0 \cos [\ell  (\theta-\Omega t)].
\end{equation}
Once mean-field interactions are included in the same way as done in
(\ref{eq:gpe}), equation (\ref{1D rotating}) can be casted in the simple form

\begin{equation} \label{1D similar}
\left[-\frac{\partial^2}{\partial \theta^2} + i \Omega
  \frac{\partial}{\partial \theta} +  V_0 \cos (\ell \theta) + {\tilde
    g} |\varphi|^2 \right]
\varphi (\theta) = {\tilde \mu} \varphi(\theta)
\end{equation}
which shows a strong formal analogy with the problem of a
Bose-Einstein condensate in a 1D optical lattice with
periodic boundary conditions ~\cite{Ueda}.
Here, $\varphi(\theta)$ is the longitudinal wavefunction, $\tilde{g}$
is the coupling constant due to interatomic interactions,
$\tilde{\mu}$ is the chemical potential shifted by constant potential
terms and all energies have been expressed in units of
$\hbar^2/(2MR^2)$.
The quantity $\Omega/2$ plays the role of the quasi-momentum for a Bloch
wave in the periodic potential of the lattice, with a subtle
difference arising from the different periodic boundary conditions: in
the periodic potential of the lattice, the allowed values of
quasi-momentum are integer multiples of $\hbar 2 \pi / Nd$, $d$ being
the lattice spacing and $N$ being the number of lattice wells present
in the system.
On the other hand, all values of $\Omega$ are allowed in the present
case of a rotating waveguide, the single-valuedness of the wavefunction
$\varphi(\theta)$ being ensured by the Bloch theorem on the whole
length of the ring, i.e. $\varphi (\theta+2\pi / \ell)=\exp [i 2 n \pi / \ell]
\varphi(\theta)$ with $n$ integer.
The complete band structure is generated by the set of
values $n=0,1, 2, \dots \ell-1$, giving rise to $\ell$ independent
sub-bands periodic in $\Omega$ with period $2\ell$.

The circulation of the different states is a function of $n$ and
$\Omega$ (for instance the lowest energy states at $2n-1<\Omega<2n+1$
have circulation $n$). The $\ell$-fold modulation of the potential
along the waveguide allows only mixing of states at circulation $n$
with states at circulation $n \pm \ell$, in correspondence of the
sub-band gaps. Hence, following adiabatically a given sub-band by very
slowly increasing (or decreasing) the rotation frequency, one can
transform a state at a given circulation $n$ into a state at
circulation $n \pm \ell$.

\section{Experimental issues}
\label{experimental}

A possible way of achieving experimentally an annular condensate with strong transverse confinement is to use a magnetic toroidal trap, as reported
in~\cite{Stamper_Kurn,Riis}.  In this section, a completely different experimental configuration is proposed, and its advantages pointed out.  For the sake
of simplicity, we shall concentrate our attention to the most relevant case of a planar waveguide whose axis lays on the horizontal $xy$ plane.  A
strong confinement in the vertical $z$ direction can be obtained using a horizontal light sheet which provides a potential $V_z(z)$, while the
curvilinear waveguide profile can be created by rapidly moving the focus point of a red-detuned laser beam (acting as an optical tweezer~\cite{tweezer})
along the curve $\mathcal{C}$, at a, possibly time-dependent speed $v(t)$. This can be obtained e.g. by reflecting the laser light onto vibrating
mirrors or using acousto-optic modulators. If the movement of the focus point is much faster than the transverse trapping frequency, then the atoms will
see the following averaged potential

\begin{equation}
V(x,y,z) = V_z (z) + \int_0^T \frac{\textrm{d}t}{T}
V_\textrm{tw}(x-x_\mathcal{C}(t),y-y_\mathcal{C}(t)),
\end{equation}
where $V_z(z)$ is the trapping potential due to the light sheet and $V_\textrm{tw}(x,y)$ the trapping potential due to the optical tweezer.  The pair
$(x_\mathcal{C}(t),y_\mathcal{C}(t))$ defines the position of the laser focus on the $xy$ plane at time $t$, which spans the curve ${\mathcal C}$ in the
period $T$ ($x$ and $y$ are here cartesian coordinates). The tweezer potential is attractive, and can be taken as having a Gaussian shape of waist $w$
and peak value $V_{\rm tw}^0<0$

\begin{equation}
V_{\rm tw}(x,y)=V_{\rm tw}^0\,\exp[-(x^2+y^2)/w^2].
\end{equation}
It is then easy to obtain expressions for the potential terms
$V_\perp$ and $V_{\rm ext}$ appearing in the Hamiltonian (\ref{schrodinger})

\begin{eqnarray}
\omega_\perp^2(s)&=&\frac{\sqrt{\pi} w}{T\,v(s)} \,
\omega_\textrm{tw}^2 \quad \textrm{with} \quad \omega_\textrm{tw}^2 =
\frac{2 |V_{\rm tw}^0 (s)|}{m  w^2}\, ,
\label{omega_perp} \\
V_{\rm ext}(s)&=& \frac{\sqrt{\pi} w}{T\,v(s)} \, V_{\rm tw}^0 (s)\label{V_ext}.
\end{eqnarray}
A remarkable fact has to be noted: both $\omega_\perp^2(s)$ and
$V_{\rm ext}(s)$ are inversely proportional to the speed of the laser
focus when passing in the neighbourhood of the point $s$ under
examination, and to its overall orbital period.  A possible way of
adding a spatially modulated external potential along the waveguide is
therefore to simply modulate the speed of motion of the focus along
the curve.

Plugging in (\ref{omega_perp}) typical values of intensity $I=100$~W
and detuning $\delta=300$~nm for optical tweezers with $w =
30$~$\mu$m, one can see that transverse extensions as low as
1.16~$\mu$m could be achieved on a 150~$\mu$m radius torus with such a
device, ensuring the validity of the strong confinement
hypothesis. The main advantage of this method over the magnetic traps
studied e.g. in~\cite{Stamper_Kurn} is that any shape of the curve
$\mathcal{C}$ can be obtained without any additional difficulty, and
this can also be modified in the course of the experiment in order to
obtain e.g. a rotating waveguide.

Another important requirement for the study of supercurrent to be
possible is that the longitudinal potentials are not strong enough to
fragment the condensate.  As an example, it is interesting to estimate
the effect of a tilting of the light sheet that is used to vertically
confine the atoms.  For a tilting angle $\alpha$ from the horizontal,
the potential energy difference at the extrema of the waveguide due to
gravity is equal to $2M g_G R \sin \alpha$, where $g_G$ is the gravity
field acceleration and $2R$ is the horizontal distance between
opposite points of the waveguide.  Within the Thomas-Fermi
approximation, fragmentation occurs if the gravitational energy
difference is larger than the mean-field interaction energy, that is
if

\begin{equation}
2M g_G R \sin \alpha > \frac{g_{1D} N_0}{2\pi R}.
\end{equation}
For a system of $N_0=10^6$ atoms of $^{87}$Rb with scattering length
$a=5.77$ nm in a waveguide of radius $R=150\,\mu$m and
$\sigma=1.16\,\mu$m, the system remain connected provided $\alpha <
0.15^\circ$, which is a rather stringent but not unreachable
experimental requirement. Note that the gravitational potential does
not rotate with the waveguide when this is set into rotation, so that
it might possibly act as a defect moving through the one-dimensional
fluid~\cite{Pavloff,Ueda2}.

In the case of very tight confinement and frozen transverse dynamics
the issue of phase fluctuations in the 1D condensate comes into play.
We believe that the rotational properties of the condensate should not
however be disturbed at least in the meanfield regime at low enough
temperature.

Another issue of great importance from the experimental point of view
is the possibility of measuring the number of vorticity quanta present
around the annular condensate.  A measurement after expansion has been
recently predicted to be able to provide a clear answer~\cite{tozzo},
but a non-destructive measurement would be preferable in view of
applications as a rotation sensor. In the case of a rotating
non-circular (e.g., elliptical) waveguide, the deformation of the
periodic density modulation due to the centrifugal potential should in
principle provide a way of measuring the supercurrent.  If this signal
is too weak to be detected, one could still resort to other
techniques, e.g. the analysis of collective modes~\cite{brian} or the
measurement of the momentum distribution by means of Bragg
spectroscopy~\cite{muniz06} or slow-light imaging~\cite{slowlight}.

\section{Conclusion}
\label{Conclu} 

In this paper, we have developed a formalism which is able to reduce
the three-dimensional problem of atomic propagation along an atomic
waveguide to one-dimensional equations. Under a strong transverse
confinement hypothesis, the transverse extension of the wavefunction
is much smaller than the curvature radius of the waveguide and the
wavefunction can be factorized into the product of its longitudinal
and transverse parts. Our formalism is then applied to a novel concept
of optical waveguide which combines the possibility of having a strong
confinement with a great flexibility in the design of the, possibly
rotating, waveguide shape.

Such a description provides a simple yet accurate starting point for
analytical studies and numerical simulations, as well as for the
design of experimental setups. 
Our framework is in fact able to considerably simplify the theoretical
analysis while still keeping track of the relevant degrees of freedom
in a quantitative way.
Possible applications range from the determination of the best
experimental sequence to nucleate a supercurrent along a  ring-shaped
Bose-Einstein condensate, to the study of the response to a global
rotation in a sort of matter-wave gyroscope.

\acknowledgements

Stimulating discussions with C. Tozzo and F. Dalfovo are warmly
acknowledged. S.Schwartz thanks P. Leboeuf, N. Pavloff, S. Richard,
J.P. Pocholle and A. Aspect for fruitful discussions. 
S.Schwartz acknowledges financial support
from European Science Foundation in the framework of the QUDEDIS program.


\begin{thebibliography}{99}
\bibitem{notation} The coupling constant $g_{3D}$ should not be
confused with the longitudinal wavefunction $g(s)$.
\end{thebibliography}
\end{document}